\documentstyle[11pt,sprocl_mod,epsfig]{article}
\def\be{\begin{equation}}
\def\ee{\end{equation}}
\def\ba{\begin{eqnarray}}
\def\ea{\end{eqnarray}}

\def\nl{\nonumber\\}
\begin{document}
\thispagestyle{empty}
%%%%%%%%%%%%%%%%%%%%%%%%%%%%%%%%%%%%%%%%%%%%%%%%%%%%%%%%%%%%%%%%%%%%%%%%
\thispagestyle{empty}
%%%\begin{flushleft}
%%%{\tt
%%%DESY 99-015
%%%\\
%%%February 1999
%%%\\
%%%hep-ph/9902408
%%%}
%%%\end{flushleft}

\bigskip

\title{%
HARD-PHOTON EMISSION IN $e^+e^-\to  {\bar f} f$ 
WITH REALISTIC CUTS
}
\author{%
P. CH. CHRISTOVA\footnote{Supported by Bulgarian foundation for
Scientific Research with grant $\Phi$--620/1996.}  
}
\address{%
Faculty of Physics, Bishop Konstantin Preslavsky Univ., 
\\ Shoumen, Bulgaria
\\ E-mail: penka@main.uni-shoumen.acad.bg
} 
\author{M. JACK, T. RIEMANN
}
\address{DESY Zeuthen, Platanenallee 6,
\\ D-15738 Zeuthen, Germany
\\ E-mails: jack@ifh.de, riemann@ifh.de
}
%==========================================================================
\maketitle\abstracts{ 
We derive compact analytical formulae of the Bonneau-Martin type for the
reaction $e^+e^- \to {\bar f} f \gamma $ with cuts on minimal energy and
acollinearity of the fermions,
where the photons may be emitted both from the initial or final states. 
Soft-photon exponentiation is also taken into account.
}
%==========================================================================
One of the cleanest scattering processes at elementary particle accelerators
is fermion-pair production in $e^+e^-$ annihilation, potentially
accompanied by one or few photons:
\begin{eqnarray}
  \label{eq:1}
 e^+e^- \to  {\bar f} f + (n)\gamma .
\end{eqnarray}
Initial-state corrections may be written as an integral over the
(normalized) invariant mass squared $R=s'/s$ of the final-state fermion pair:
\begin{eqnarray}
  \label{eq:2}
 \sigma_T^{ini}(s) = \int dR~ \sigma^0(s')~ \rho_T^{ini}(R),  
\end{eqnarray}
where $\sigma^0(s')$ is an effective Born cross-section.
The radiator function $\rho_T^{ini}(R)$
for the initial-state first-order corrections 
to the total cross-section $\sigma_T$,  
with soft-photon exponentiation, is \cite{Greco:1975rm}: 
\begin{eqnarray}
  \label{eq:3}
  \rho_T^{ini}(R) &=& \left(1+{\bar S}\right)\beta (1-R)^{\beta-1} 
   + {\bar H}_{T}^{ini}(R),
\ea
with
\ba
  \label{eq:4}
         {\bar S} &=& \frac{3}{4}\beta +\frac{\alpha}{\pi}Q_e^2
\left(\frac{\pi^2}{3} - \frac{1}{2}\right) + h.o.,~~~~~ 
  \beta = \frac{2\alpha}{\pi} Q_e^2 \left( \ln \frac{s}{m_e^2}-1\right) ,
\ea
and
\ba
  \label{eq:3a}
 {\bar H}_{T}^{ini}(R)&=&\left[ H_{BM}(R)-\frac{\beta}{1-R}\right] +h.o. ,
\end{eqnarray}
where $h.o.$ stands for higher orders, and
\cite{Bonneau:1971}:
\begin{eqnarray}
  \label{eq:bm}
  H_{BM}(R) = \frac{1}{2}~\frac{1+R^2}{1-R}~\beta.
\end{eqnarray}
Experiments at LEP1, SLC, LEP2, and those planned at a linear collider aim at
precisions well below a per cent and need theoretical predictions with an
accuracy of the order of 0.1~\% or better.  
A basic ingredient of the predictions is the complete photonic $O(\alpha)$
correction including initial and final state radiations and 
their interferences:
\ba
\label{sigma}
\sigma(s) &=& \sigma^0(s) + \sigma^{ini}(s) + \sigma^{int}(s) +
\sigma^{fin}(s) .
\ea
These corrections have to be determined for two basic quantities: The total
cross-section $\sigma_T(s)$ and
the forward-backward asymmetry
$A_{FB} = \sigma_{F-B}/\sigma_T$; other asymmetries may then easily be
derived. 

Basically, there are two experimental set-up's to be treated:
\begin{itemize}
\item[(i)] a lower cut on $s'$, $s'_{min}\geq 4m_f^2$, often applied to
  quark-pair production;   
\item[(ii)] combined cuts on acollinearity $\xi$, $\xi_{max}\leq 180^{\circ}$,
  and minimal energy $E_{min}$, $E_{min}\geq m_f$, of the fermions; often
  applied to lepton-pair production.  
\end{itemize}
Both cut settings may be combined with an acceptance cut $c$, $c\leq1$, on the
cosine of the fermionic production angle $\cos\theta$.  
 
For case (i), a generalization of the Bonneau-Martin formula, including the
complete first-order photonic corrections together with soft-photon
exponentiation, may be found in \cite{Bardin:1989cw} without and in  
\cite{Bardin:1991de} with acceptance cut.
The extremely compact expressions for case (i) with $c=1$ get quite more
involved when the acceptance cut is applied. 

\bigskip

In this article, we give the complete first-order photonic corrections
together with soft-photon exponentiation for case (ii).

\begin{figure}[tbhp]
  \begin{center}
\vspace*{-1.2cm}
%--- 
  \mbox{%
  \epsfig{file=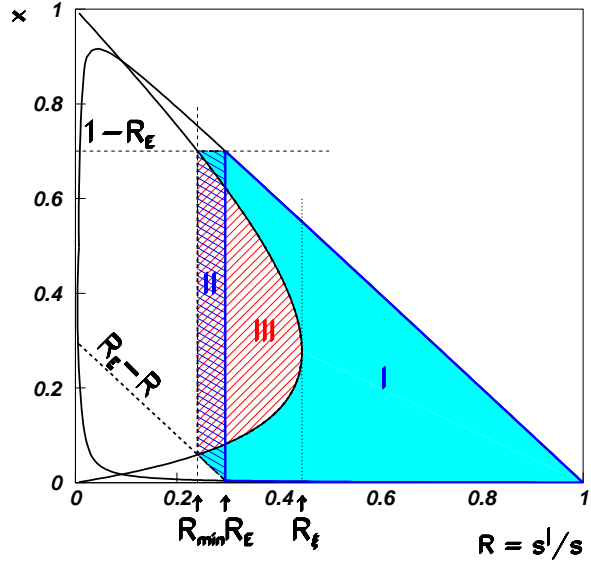
          ,height=9cm  % this is the height of the figure (optional)
          ,width=9cm   % this is the width of the figure (optional)
         }%
  }
\caption[]{\label{dalitz} 
 Phase space with cuts on acollinearity and (lower) fermion energy. 
}
    \label{fig:1}
  \end{center}
\end{figure}

A three-fold analytical integration of the squared matrix elements
is performed in order to get the integrand
of the last one, the $s'$-integration over $R$, which is assumed to be
performed  numerically. 
One may use the phase-space parameterization
derived in \cite{Passarino:1982zp}.
The kinematical regions of two variables are shown in figure \ref{fig:1}:
the (normalized) invariant mass $x$ of (photon + anti-fermion) in their rest
system, and $R$ as introduced above. 
The first and third analytical
integrations are over the full production angle of the photon,
$\phi_{\gamma}$, in the (photon + fermion) rest system, and over the
cosine of the anti-fermion in the center-of-mass system, $\cos\vartheta$.
Both angles are completely independent of $x$ and $R$.
As figure \ref{fig:1} shows, we have to determine radiators $\rho_B^b$ 
($B=T,F$--$B$ and $b=ini,int,fin$) in three phase-space regions:  
\ba
\sigma_B^{b}(s) = \left[ \int_{\mathrm{I}} + \int_{\mathrm{II}}  
- \int_{\mathrm{III}} \right]~ d\phi_{\gamma} ~ dx ~ ds' ~  d\cos\vartheta 
\frac{d\sigma_B^b(A)}{d\phi_{\gamma}dx ds' d\cos\vartheta} ,
\label{sig}
\ea
Region I applies to the simple $s'$-cut.
%%%% 24-02-99: integration boundaries for R and x explicitely added: 
The integration over $R$ extends from $R_{min}$ to 1,
\ba
\label{rmin}
R_{min} &=& R_E \left(1 - \frac{\sin^2(\xi_{max}/2)}
{1-R_E\cos^2(\xi_{max}/2)} \right) . 
\ea
In each of the three regions, 
the boundaries for the integration over $x$ are, for a given value of
$R$:
\ba
\label{ibound}
x_{max,min} &=& \frac{1}{2} (1-R)~ (1 \pm A),
\ea

%---------------------

\noindent
where the parameter $A=A(R)$ depends in every region on only one of
the cuts applied:  
%---------------------
\ba
A_{\mathrm{I}} &=& \sqrt{1-\frac{R_{m}}{R}},
\label{A1}
\\
A_{\mathrm{II}} &=& \frac{1+R-2R_E}{1-R}, 
\label{A2}
\\
A_{\mathrm{III}} &=& \sqrt{1 - \frac{R(1-R_{\xi})^2}{R_{\xi}(1-R)^2}}, 
\label{A3}
\ea
%---------------------

with:
\begin{eqnarray}
  \label{eq:rs}
R_{m}  &=& \frac{4m_f^2}{s},
\\
  \label{eq:re}
R_E &=& \frac{2E_{min}}{\sqrt{s}}, 
\\
  \label{eq:rx}
R_{\xi} &=& \frac{1-\sin(\xi_{max}/2)}{1+\sin(\xi_{max}/2)}.  
\end{eqnarray}
%====================================================================
\section*{ Initial State Radiation }
\label{sec:ini}
%====================================================================
Here, for $\sigma_T^{ini}(s)$ the Bonneau-Martin function gets replaced by:
%------------------
\ba
\label{sigtA}
H_T^{ini}(R,A) &=& 
\frac{3\alpha}{4\pi} Q_e^2 
\left[
\left(A+\frac{A^3}{3}\right) \frac{1+R^2}{1-R}
\left( \ln\frac{s}{m_e^2}-1\right) + (A-A^3) \frac{2R}{1-R}
\right].
\ea
%------------------
In $\sigma_{F-B}^{ini}(s)$, the corresponding hard radiator part is:
%------------------------------------------------------------------ 
\begin{eqnarray}
\label{inifb1}
H^{ini}_{F-B}(R;A\ge A_0)&=&
\frac{\alpha}{\pi} Q_e^2
\Biggl\{
\frac{1+R^2}{1-R}\Biggl[
\frac{4 R}{(1+R)^2} \left( \ln\frac{s(1+R)^2}{4m_e^2R} - 1 \right)
\nonumber\\
&&
%%% 12-03-99: superflous doubly factors eliminated; Fortran was ok. 
% -~\frac{1+R^2}{(1+R)^2(1-R)}
  -~\frac{1}{(1+R)^2}
\left[
y_+ y_- \ln\left|y_+ y_-\right| 
%%% 24-02-99: plus sign added; was a typo:
+
{4 R}\ln(4 R)
\right]
\nonumber\\
&&-~(1-A^2) \left( \ln\frac{s}{4m_e^2(1+A)^2R} - 1\right)
\Biggr]
+\frac{4A(1-A)R}{1-R}
\Biggr\}, 
\\%==============================================
\label{radinih_fb_a} 
H^{ini}_{F-B}(R;A<A_0)&=&
\frac{\alpha}{\pi} Q_e^2
\Biggl\{
\frac{1+R^2}{1-R}
\left[
-\frac{y_+ y_-}{(1+R)^2}
\ln\left|\frac{y_+}{y_-}\right|
+(1-A^2)\ln\frac{1+A}{1-A}\right]
\nonumber\\
&&
+~\frac{8 A R^2}{(1+R)(1-R)}
\Biggr\}.
%------------------------------------------------------------------ 
\end{eqnarray}
%------------------------------------------------------------------ 
These functions have to be used in (\ref{eq:3}) and its analogue, with 
equal ${\bar S}$, for $\sigma_{F-B}^{ini}$.
The following definitions are used:
\begin{eqnarray}
  \label{eq:a0}
A_0 &=& \frac{1-R}{1+R},
\\
  \label{eq:yy}
y_{\pm} &=& (1-R) \pm A(1+R).
\end{eqnarray}
In region I, the above expressions  (\ref{sigtA}) and (\ref{inifb1})
reduce to those known from \cite{Bonneau:1971} and \cite{Bardin:1989cw}.
In this region the radiators diverge for $R \to 1$, and soft-photon
exponentiation and the subtraction
$\beta/(1-R)$ is applied there (and only there) in order to get ${\bar 
H}_B^{ini}(R)$; see (\ref{eq:3a}).
%====================================================================
\section*{ Initial-Final  State Interferences}
\label{sec:int}
%====================================================================
In the initial-final state interferences,
the effective Born cross-sections depend on both $s$ and $s'$
as well as on the type of exchanged vector particles $V_i$ (e.g. photon and or
$Z$):
\ba
\label{intro}
\sigma_B^{int}(s) &=& \int dR \sum_{V_i,V_j=\gamma, Z} 
\sigma_{\bar B}^0(s,s',i,j)~ \rho_B^{int}(R,A,i,j).
\ea
For $B$=$T$ it is ${\bar B}$=$F$--$B$ and vice versa.
We give as examples simple model-independent Born expressions in
order to fix the overall normalization:
\ba
\label{effbornt}
\sigma_T^0(s,s') &=& \sum_{V_i,V_j=\gamma, Z} 
\sigma_{T}^0(s,s',i,j)
=\frac{4\pi\alpha^2}{3s'}~ {\cal V},
\\
\label{effbornfb}
\sigma_{F-B}^0(s,s') &=& \sum_{V_i,V_j=\gamma, Z} 
\sigma_{F-B}^0(s,s',i,j) =
\frac{\pi\alpha^2}{s'}~ {\cal A},
\ea
with
\ba
{\cal V} &=&  
Q_e^2Q_f^2 +   |Q_eQ_f|v_ev_f \Re e [\chi(s)+\chi(s')] 
+ (v_e^2+a_e^2)(v_f^2+a_f^2)
\Re e[\chi(s)\chi^*(s')], 
\\
{\cal A} 
&=&
 |Q_eQ_f|a_ea_f \Re e [\chi(s)+\chi(s')] +
4v_ea_ev_fa_f  \Re e[\chi(s)\chi^*(s')] ,
\ea
and with the following $Z$ boson propagators: 
\ba
\label{propz}
\chi(s) &=& \frac{G_{\mu}M_Z^2}{8\sqrt{2}\pi\alpha}
\frac{s}{s-M_Z^2+iM_Z\Gamma_Z}.
\ea
The radiator functions are:
\begin{eqnarray}
\label{eq:rhoint}
\rho_B^{int}(R,A;i,j) &=& \delta(1-R) \left[S_B + b_B(i,j) \right] 
+ \theta(1-R-\epsilon) H_{B}^{int}(R,A).
\end{eqnarray}
The box contributions  $b_T(i,j)$ may be taken from equations (116)
and (118) (to be multiplied by 4/3) of
\cite{Bardin:1991fu} and the $b_{F-B}(i,j)$ from equations (123) and
(126).
 For convenience, we give the soft corrections explicitly:
 \begin{eqnarray}
   \label{eq:intsoftt}
S_T^{int}  &=& 8 \frac{\alpha}{\pi} Q_eQ_f
\left( 1-\ln\frac{2\epsilon}{\lambda}\right),
\\
     \label{eq:intsoftfb} 
S_{F-B}^{int} &=& \frac{\alpha}{\pi} Q_eQ_f
\left[
-\left(1+8\ln 2\right) \ln\frac{2\epsilon}{\lambda} +4\ln^2 2 + \ln 2 +
\frac{1}{2} + \frac{1}{3}\pi^2    \right] .
 \end{eqnarray}
Finally, the hard radiator parts are:
\begin{eqnarray}
  \label{eq:intht}
  H^{int}_{T}(R,A) &=&  
-\frac{\alpha}{\pi} Q_eQ_f
\frac{4 A R (1+R)}{1-R},
\end{eqnarray}
and 
\begin{eqnarray}
  \label{eq:inthfb1}
 H^{int}_{F-B}(R,A\geq A_0)&=&
\frac{\alpha}{\pi} Q_eQ_f 
\Biggl\{
\frac{3R}{2}\left[\ln\frac{z_+}{z_-}
+\frac{2-R+\frac{5}{3}R^2}{1-R}\ln R 
\right]
\nonumber\\
&&-~\frac{1+R}{2(1-R)}
(5-2 R+5 R^2)
\ln\frac{(1+R)(1+A)}{2} 
\nonumber\\
&&+~\frac{1}{4(1-R)}\Biggl[
\frac{(1-4R+R^2)[A(1+R)^2-(1-R)^2]}{1+R}
\nl &&+~ 2 A (1-A) (1+R^3)\Biggr]
\Biggr\},
\\
  \label{eq:inthfb2}
H^{int}_{F-B}(R,A<A_0)&=&
\frac{3\alpha}{2\pi} Q_eQ_f ~ R
\Biggl\{
\ln\frac{z_+}{z_-}
-
\frac{2-R+\frac{5}{3}R^2}{1-R}
%\frac{1}{1-R}\left(2-R+\frac{5}{3}R^2\right)
\ln\frac{1+A}{1-A} 
+ A (1-R)  
 \Biggr\},
\end{eqnarray}
with
\ba
  \label{eq:zz}
z_{\pm} &=& (1+R) \pm A(1-R).
\ea
Again, for $A\to 1$ the $H^{int}_{T}(R,A)$ and $H^{int}_{F-B}(R,A\geq A_0)$
approach the known expressions of the $s'$ cut given in
\cite{Bardin:1989cw}\footnote{
We realized a misprint in eq. (22) of \cite{Bardin:1989cw}; the
non-logarithmic terms have to be multiplied there by $1/(1+R)$.   
}.  
%====================================================================
\section*{Final State Radiation}
\label{sec:fin}
%====================================================================
The final state corrections to order $O(\alpha)$ are:
\begin{eqnarray}
\label{fins}
\sigma_B^{fin}(s) &=& \sigma_B^0(s) \int dR~ \rho_B^{fin}(R,A),
\ea
with
\ba
  \label{eq:f1}
\rho_B^{fin}(R,A) &=& \delta(1-R) S_f + \theta(1-R-\epsilon)
H_{B}^{fin}(R,s,A), 
\\
S_f &=& {\bar S}_f + \beta_f \ln \epsilon,
\ea
where ${\bar S}_f$ and $\beta_f$ are the final state's analogues of 
${\bar S}$ and $\beta$.
The hard radiators are:
\begin{eqnarray}
  \label{eq:fint}
H^{fin}_{T}(R,s,A)&=& 
\frac{\alpha}{\pi} Q_f^2
\left[
\frac{1+R^2}{1-R} \ln\frac{1+A}{1-A}
-\frac{8 A m_f^2/s}{(1-A^2)(1-R)}-A(1-R)
\right],
\\
%------------------------------------------------------------------ 
  \label{eq:finfb}
H^{fin}_{F-B}(R,s,A)
&=& 
H^{fin}_{T}(R,A)
+
\frac{\alpha}{\pi} Q_f^2
\left[
A(1-R)
-(1+R)\ln\frac{z_+}{z_-}
\right].
%------------------------------------------------------------------  
\end{eqnarray}
Common initial- and final state soft-photon exponentiation may be
performed as follows \cite{Nicrosini:1988sw,Bardin:1991fu}:
\begin{eqnarray}
  \label{eq:comex}
 \sigma_B^{ini+fin}(s) &=& 
\int dR ~\sigma^0(s') ~\rho_B^{ini}(R,A) ~{\bar \rho}_B^{fin}(R,s',A), 
\end{eqnarray}
with 
%%% 24-02-99: next formula slightly rewritten, the original was also
%%% ok but less explicit with the exact definition of soft exponentiation
%%% now also more explicit that only region I, R>R_E, is involved in SPE
\begin{eqnarray}
  \label{eq:rbar}
  {\bar \rho}_B^{fin}(R,s',A)  = 
%\int_{R_{min}/R}^{1} du ~\rho_B^{fin}(u,s',A').
(1-R_E)^{\beta'_f}(1+S'_f) 
+ \int_{R_{min}/R}^{1} du \left[H^{fin}_{B}(u,s',A')-\frac{\beta'_f}{1-u}
\theta(R-R_E)\right].
\end{eqnarray}
The soft part of $\rho_B^{fin}(u,s',A')$, $A'=A(u)$, and $\beta'_f$
are derived from 
(\ref{eq:3}) by replacing there $Q_e$ by $Q_f$ and $s/m_e^2$ by
$s'/m_f^2$.
%%% 24-02-99 while the hard 
%%% 24-02-99radiators are:
%%% 24-02-99\begin{eqnarray}
%%% 24-02-99   \label{eq:fin6}
%%% 24-02-99{\bar H}_{B}^{fin}(u,s',A') &=& H^{fin}_{B}(u,s',A') 
%%% 24-02-99- \delta_{A',1} \frac{\beta'_f}{1-u}.
%%% 24-02-99\end{eqnarray}
%%% 24-02-99The subtraction $\beta'_f/(1-u)$ is, of course, only taken
%%% 24-02-99in region I  
%%% 24-02-99where the soft-photon exponentiation is done (and $A' \approx 1$). 
We mention here that in the hard radiators the integration over
$u$ may also be performed analytically. 
In region III, one has to interchange for this the order of
integration over $u$ and $x$ \cite{Montagna:1993mf}. 
%====================================================================
\section*{Conclusions}
\label{sec:sum}
%====================================================================
We recalculated the photonic corrections with acollinearity cut having 
applications in the Fortan program {\tt ZFITTER} in mind  \cite{zfitter}.   
When the code was created in 1989
\cite{MBilenky:1989ab}, an accuracy of 0.5 \% at LEP1 was assumed
to be needed
\footnote{
For other recent comparisons see e.g. \cite{Christova:1998tc,PCP:99} and
for the influence of higher order corrections e.g. \cite{PCP:99}, and
references therein.
}.

We performed several numerical applications of the above formulae.
For this purpose, the package {\tt acol.f} was added.
As a result, we conclude that {\tt ZFITTER} until version 5.20 treats
the $O(\alpha)$ photonic corrections with acollinearity 
cut with a numerical accuracy for $\sigma_T$ of not less than about
0.4 \% near the $Z$ resonance (LEP1 energy region, $\sqrt{s}$ within
$M_Z \pm 3$ GeV)  or better (at resonance). 
The coding with {\tt acol.f} gave a numerical agreement of
$\sigma_T$ for leptons, 
with $\xi_{max}\leq 10^{\circ}$ and $E_{min}=1$ GeV, with predictions 
from {\tt TOPAZ0} v.4.4 \cite{topaz0} at LEP1 of 0.01 \% (at the
wings) or better (at resonance) \cite{Bardin:1999oo}.
For $A_{FB}$, we estimate at LEP1 energies the 
accuracy of the 
%%% 24-02-99: to be more precise,  $O(\alpha)$ added
 $O(\alpha)$
photonic corrections with acollinearity cut in {\tt
ZFITTER} until v.5.20 
to be about 0.02 \% or better at the resonance and  
about 0.13 \% or better at the wings.  
The numerical limitations at LEP1  
are due to the initial-final state interference.

For applications at higher energies, the accuracy of {\tt ZFITTER}
with acollinearity cut was not dedicatedly 
controlled until recently, although there was reason to suspect that it comes
out much worse than at LEP1. 
With the cuts mentioned,  
the accuracy of v.5.20 at LEP2 is again limited by the initial-final
interference but not less than roughly 1 \%.

A higher accuracy in the acollinearity mode is prevented in {\tt ZFITTER}
until v.5.20 by several reasons. 
The main reason is a neglect of a certain class of ordinary, angular
dependent  
$O(\alpha)$ terms in the initial state and in the initial-final state
interference hard radiator parts. 
Some numerical approximations in the treatment of final state
radiation may also be influential; this has not been studied in
detail so far.

Finally we should like to mention that the correct $O(\alpha)$ hard
radiators for the angular distributions 
and for the integrated cross-sections with 
%%% 24-02-99: to be more precise:
cuts on both acollinearity and angular acceptance
%finite acceptance cuts 
also have been determined within this project and will be published
elsewhere together with a scetch of the calculations and more
numerical results. 
%==========================================================================
\subsection*{Acknowledgments}
%--------------------------------------------------------------------------
We would like to thank D.~Bardin, S.~Vasileva, M.~Gr\"unewald, G.~Passarino,
and S.~Riemann for numerical comparisons, and D. Bardin for a careful 
discussion of the conclusions. 
%==========================================================================

\bigskip

\section*{References}
%======================================================================= 
%\def\href#1#2{#2} 
%\bibliography{%
%/home/hades/jack/Radcor98/proceed/text/radcorr%
%/home/phoenix/riemann/Bibliography/radcorr% 
%/home/tord/Tex/Bib/radcorr% 
%} 
%\bibliographystyle{%
%/home/hades/jack/Radcor98/proceed/text/utphys_t%
%/home/phoenix/riemann/Bibliography/utphys_t%
%/home/tord/Tex/Bib/utphys_t%
%} 

%======================================================================= 
\end{document}